\documentclass[twocolumn]{article}
\usepackage{amsfonts}
\usepackage{epsfig}
\usepackage{dcolumn}
\begin{document}
\title{Scale-free Segregation in Transport Networks}

\author{Ph. Blanchard, D. Volchenkov
\\
{\small  Bielefeld-Bonn Stochastic Research Center (BiBoS)},\\
{\small Bielefeld University, Postfach 100131, 33501, Bielefeld, Germany}\\
{\small Email: volchenk@physik.uni-bielefeld.de}
}
\date{\today}

\twocolumn[
\maketitle
\begin{abstract}
Every route of a transport network approaching equilibrium can be
represented by a vector of Euclidean space which length quantifies
its segregation from the rest of the graph. We have empirically
observed that the distribution of lengths over the edge
connectivity in many transport networks exhibits scaling
invariance phenomenon. We give an example of the canal network of Veneice
to demonstrate our result. The method is applicable to any transport network.
\end{abstract}

\vspace{0.2cm}

\leftline{PACS: 89.65.Lm, 89.75.Fb, 05.40.Fb, 02.10.Ox }

\vspace{0.5cm}
]

Transport networks are used
to model the flow of commodity,
 information, viruses, opinions, or traffic.
They typically represent the
 networks of roads, streets, pipes,
aqueducts, power lines, or
 nearly any structure
 which permits either vehicular
movement or flow of some commodity,
  products, goods or service.
The major aim of the analysis is to
determine the structure and properties
 of
 transport networks that are important
for  the emergence of
 complex flow patterns
of vehicles
(or pedestrians) through the network
such as the  Braess paradox \cite{Braess}.
This counter-intuitive
phenomenon occurs when
adding more resources to a transportation network
(say, a new road or a bridge)
worsens the quality of traffic by creating
longer delays for the drivers,
rather than alleviate it. The Braess paradox
has been
observed in the street
 traffic of New York City and Stuttgart, \cite{Kolata}.

In the present Letter, we show that while approaching equilibrium,
a transport network can be embedded into Euclidean space 
$\mathbb{R}^{N-1}$, $N$ being a number of vertices.
Then, every edge of the network is
represented by a vector which length quantifies its segregation
from the rest of the graph. We have empirically observed that the
distribution of lengths over the edge connectivity in urban
transport networks exhibits scaling invariance
phenomenon.  The relation between the connectivity of city spaces
and their centrality known as {\it intelligibility} is a key
determinant of human behaviors in urban environments,
\cite{Hillier:1999}.

In most of researches
devoted to the improving  of transport networks,
a  {\it primary} graph
  representation of urban networks is used in which
streets and routes are considered as edges
  of a planar graph, while the traffic end points and street
  junctions are treated as nodes.
The usual
   city map based on Euclidean geometry can be considered as an
   example of primary city graphs.

However, another graph representation can be useful if we are interested
in describing the transport network at equilibrium.
Given a connected undirected graph $G(V,E)$, in which $V$ is the
set of nodes and $E$ is the set of edges, we introduce the traffic
function $f: E\to(0,\infty[$ through every edge $e\in E$. It then follows
  from the Perron-Frobenius theorem \cite{PerronFrobenius} that the linear equation
\begin{equation}
\label{Lim_equilibrium}
f(e)\,=\, \sum_{e'\sim\, e}\,f(e')\,\exp\left(\,-h\,\ell\left(e'\right)\,\right),
\end{equation}
where the sum is taken over all edges $e'\in E$ which have a common node with $e$,
has a unique positive solution $f(e)>0$, for every edge $e\in E$, for a
fixed positive constant $h>0$ and a chosen set of positive  {\it metric
 length} distances $\ell(e)>0$. This solution is naturally identified
 with the traffic equilibrium state of the transport network defined on
  $G$, in which the permeability of edges depends upon their lengths.
The parameter $h$  is called the volume entropy of the graph $G$, while
 the volume of $G$ is defined as the sum
$
\mathrm{Vol}(G)=\frac 12\sum_{e\in E}\ell(e).
$
The volume entropy $h$ is defined to be the exponential growth of the balls in
a universal covering tree of $G$ with the lifted metric, \cite{Manning}-\cite{Lim:2005}.

The degree of a node $i\in V$ is the number of its neighbors in
$G$, $\deg_G (i)=k_i$. It has been shown in \cite{Lim:2005} that
among all undirected connected graphs of normalized volume,
$\mathrm{Vol}(G)=1$, which are not cycles and for which $k_i\ne 1$ for all
nodes,
 the minimal  value of the volume entropy,
$\min(h)=\frac 12\sum_{i\in V}k_i\,\log\left(k_i-1\right)$  is attained
for the length distances
\begin{equation}
\label{ell_min}
\ell(e)\,=\,\frac {\log\left(\left(k_i-1\right)
\left(k_j-1\right)\right)}{2\,\min(h)},
\end{equation}
where $k_i$ and $k_j$ are the degrees of the nodes linked by $e \in E$.
It is then obvious that substituting (\ref{ell_min})
and $\min(h)$ into (\ref{Lim_equilibrium}) the
operator $\exp\left(-h \ell(e')\right)$ is given by
 a symmetric Markov transition operator,
\begin{equation}
\label{Markov_transition}
f(e)\,=\, \sum_{e'\,\sim\, e}\,\frac{f(e')}{\sqrt{\left(k_{i}-1\right)
\left(k_{j}-1\right)}},
\end{equation}
where $i$ and $j$ are the  nodes linked by $e' \in E$, and the sum in
(\ref{Markov_transition}) is taken over all
edges $e'\in E$ which share a node with $e$.
The symmetric operator (\ref{Markov_transition})
 rather describes time reversible random walks over edges than over
nodes.
In other words, we are invited to consider random walks described
by the symmetric operator defined on the {\it dual} graph $G^\star$.

The Markov process (\ref{Markov_transition}) represents the
conservation of the traffic volume through the transport network,
while other solutions of (\ref{Lim_equilibrium}) (with $h>\min(h)$) are related to
the possible termination of travels along edges. If we denote the
number of neighbor edges the edge $e\in E$ has in the dual graph
$G^\star$ as $q_e=\deg_{G^\star}(e)$, then the simple substitution
shows that $w(e)=\sqrt{q_e}$ defines an eigenvector of the
symmetric Markov transition operator defined over the edges $E$
with eigenvalue 1. This eigenvector is positive and being properly
normalized determines the relative traffic volume through $e\in E$
at equilibrium.
Eq.(\ref{Markov_transition})
relates the equilibrium transport flows on the graph $G$
to the stationary distribution of random walks defined on its dual counterpart $G^\star$
and emphasizes that
 the degrees of nodes are a key determinant of the transport networks properties.

The notion of traffic equilibrium had been introduced by J.G. Wardrop in
 \cite{Wardrop:1952} and then generalized in \cite{Beckmann:1956} to a
 fundamental concept of   network equilibrium. Wardrop's traffic
 equilibrium  is strongly tied to the human apprehension of space since
  it is required that all travellers have enough knowledge of the
  transport network they use. Dual city graphs are
 extensively investigated within the concept of
{\it space syntax}, a theory developed in the late 1970s, that seeks
to reveal the effect of spatial configurations on
the human perception of places and behavior in urban environments,
\cite{Hillier:1984,Hillier:1999}.
Spatial perception
 that shapes peoples understanding of how
a place is organized  determines eventually the pattern of local movement,
 which is quantified
by the space syntax
measure  being nothing
else, but an element of a transition probability  matrix
of a Markov chain \cite{Batty},
 with surprising accuracy \cite{Penn:2001}.
Random walks embed connected undirected graphs into the Euclidean
space $\mathbb{R}^{N-1}$. This embedding  can be used in order to compare
 nodes with respect to the quality of paths they provide for random walkers
and to construct
 the optimal coarse-graining
representations.

While analyzing a graph, whether it is primary or dual, we assign
the absolute scores to all nodes based on their properties with
respect to a transport process defined on that. Indeed, the nodes
of $G(V,E)$ can be weighted with respect to any  measure
 $m=\sum_{i\in V} m_i {\bf 1}_{i},$ specified by a set of positive numbers $m_i> 0$.
 The space $\ell^2(m)$ of square-assumable functions with respect to the
   measure $m$ is the Hilbert space $\mathcal{H}(V)$.
Among all measures which can be defined on $V$, the set of
normalized measures (or {\it densities}),
$1=\sum_{i\in V}\pi_i{\bf 1}_{i},$
 are of essential interest since they express the conservation of a
  quantity, and therefore may be relevant to a physical process.

The fundamental physical process defined on a graph is generated by the
subset of its linear automorphisms which share the property of  probability
conservation; it can be naturally interpreted as random walks.
The linear automorphisms of the graph are specified by the symmetric group
 $\mathbb{S}_N$ including all admissible permutations $p\in \mathbb{S}_N$
taking $i\in V$ to $p(i)\in V$ and preserving all of its structure.

 Markov's operators on Hilbert space form the natural language of transport networks theory.
Being defined on a connected aperiodic graph, the transition matrix of random walks $T_{ij}$
is a real positive stochastic matrix, and therefore, in accordance to the
 Perron-Frobenius theorem \cite{PerronFrobenius}, its maximal
 eigenvalue is 1, and it is simple. The left eigenvector $\pi T=\pi$
 associated with the eigenvalue 1 is interpreted as a unique
 equilibrium state $\pi$ (the stationary distribution of random walks).
The Markov operator $\widehat{T}$
is self-adjoint with respect to the normalized measure associated
to the stationary distribution of random walks $\pi$,
\begin{equation}
\label{s_a_analogue}
\widehat{T}=\frac 12
\left( \pi^{1/2} T
\pi^{-1/2}+\pi^{-1/2} T^\top
 \pi^{1/2}\right),
\end{equation}
where $T^\top$ is the transposed operator. In the theory of
  random walks  defined on graphs \cite{Lovasz} and in spectral
  graph theory \cite{Chung:1997}, basic properties of graphs are studied in
   connection with the  eigenvalues and eigenvectors of self-adjoint operators
   defined on them.
The orthonormal ordered set of real
 eigenvectors $\psi_i$, $i=1\ldots N$, of the symmetric operator $\widehat{T}$
 defines a basis in  $\mathcal{H}(V)$.

In particular, the symmetric transition operator  $\widehat{T}$ of the
random walk  defined on connected    undirected graphs is $\widehat{T_{ij}}=1/\sqrt{k_ik_j}$,
if $i\sim j$, and zero otherwise.
Its first eigenvector
    $\psi_1$ belonging to the largest eigenvalue $\mu_1=1$,
\begin{equation}
\label{psi_1}
\psi_1
\,\widehat{ T}\, =\,
\psi_1,
\quad \psi_{1,i}^2\,=\,\pi_i,
\end{equation}
describes the {\it local} property of nodes (connectivity),
since the stationary distribution of random walks  is
$
\pi_i=k_i/2M
$ where $2M=\sum_{i\in V} k_i$.
The remaining eigenvectors,
 $\left\{\,\psi_s\,\right\}_{s=2}^N$, belonging to the eigenvalues
  $1>\mu_2\geq\ldots\mu_N\geq -1$ describe the {\it global} connectedness of the graph.
For example, the eigenvector corresponding to the second eigenvalue $\mu_2$
is used to define  the spectral bisection of graphs; it is called the
Fiedler vector if related to the Laplacian matrix of a graph \cite{Chung:1997}.

Markov's symmetric transition operator $\widehat{T}$  defines a projection
 of any density $\sigma\in \mathcal{H}(V)$ on the eigenvector $\psi_1$ of the
  stationary distribution $\pi$,
\begin{equation}
\label{project}
\sigma\,\widehat{T}\,
=\,\psi_1 + \sigma_\bot\,\widehat{T},\quad \sigma_\bot\,=\,\sigma-\psi_1,
\end{equation}
in which $\sigma_{\bot}$ is the vector belonging to the orthogonal complement of
$\psi_1$.
In space syntax, we are interested in a comparison between the densities  with respect
 to random walks defined on the graph $G$.  Since all components $\psi_{1,i}>0$,
it is convenient to rescale the density $\sigma$ by dividing its
components by the components of $\psi_1$,
\begin{equation}
\label{rescaling}
\widetilde{\sigma_i} =\frac{\sigma_i}{\psi_{1,i}}.
\end{equation}
Thus, it is clear that any two rescaled densities
$\widetilde{\sigma},\widetilde{\rho}\in\mathcal{H}(V)$ differ
 with respect to random walks only by their dynamical components,
$
\left(\widetilde{\sigma}-\widetilde{\rho}\right)
 \widehat{T}^t=\left(\widetilde{\sigma}_\bot -\widetilde{\rho}_\bot\right)
\widehat{T}^t,
$
 for all $t>0$.
Therefore, we can define the
distance  $\|\ldots\|_T$ between any two densities established by random walks by
\begin{equation}
\label{distance}
\left\|\sigma-\rho\right\|^2_T =
 \sum_{t\geq 0} \left\langle \widetilde{\sigma}_\bot -\widetilde{\rho}_\bot\left|\widehat{T}^t
\right|\, \widetilde{\sigma}_\bot -\widetilde{\rho}_\bot\,\right\rangle,
\end{equation}
 or, using the spectral
representation of $\widehat{T}$,
\begin{equation}
\label{spectral_dist}
\left\|\sigma-\rho\right\|^2_T
=\sum_{s=2}^N\,\frac{\left\langle\, \widetilde{\sigma}_\bot -\widetilde{\rho}_\bot\,|
\, \psi_s\,\right\rangle\!\left\langle\, \psi_s\,
| \,\widetilde{\sigma}_\bot -\widetilde{\rho}_\bot\,\right\rangle}{\,1\,-\,\mu_s\,},
\end{equation}
where we have used  Dirac's bra-ket notations especially
convenient for working with inner products and
rank-one
operators in Hilbert space.

If we introduce a new inner product for
densities $\sigma,\rho \in\mathcal{H}(V)$
by
\begin{equation}
\label{inner-product}
\left(\,\sigma,\rho\,\right)_{T}
\,= \, \sum_{t\,\geq\, 0}\, \sum_{s=2}^N
\,\frac{\,\left\langle\,  \widetilde{\sigma}_\bot\,|\,\psi_s\,\right\rangle\!
\left\langle\,\psi_s\,|\, \widetilde{\rho}_\bot \right\rangle}{\,1\,-\,\mu_s\,},
\end{equation}
then (\ref{spectral_dist}) is nothing else but
$
\left\|\,\sigma-\rho\,\right\|^2_T\, =
\left\|\,\sigma\,\right\|^2_T +
\left\|\,\rho\,\right\|^2_T  -
2 \left(\,\sigma,\rho\,\right)_T,
$
 where
\begin{equation}
\label{sqaured_norm}
\left\|\, \sigma\,\right\|^2_T\,=\,
\,\sum_{s=2}^N \,\frac{\left\langle\,  \widetilde{\sigma}_\bot\,|\,\psi_s\,\right\rangle\!
\left\langle\,\psi_s\,|\, \widetilde{\sigma}_\bot\, \right\rangle}{\,1\,-\,\mu_s\,}
\end{equation}
is the square
of the norm of  $\sigma\,\in\, \mathcal{H}(V)$ with respect to
random walks defined on the graph $G$.

We finish the description of the Euclidean
space structure of $G$ induced by
  random walks by mentioning that
given two densities $\sigma,\rho\,\in\, \mathcal{H}(V),$ the
angle between them can be introduced in the standard way,
\begin{equation}
\label{angle}
\cos \,\angle \left(\rho,\sigma\right)=
\frac{\,\left(\,\sigma,\rho\,\right)_T\,}
{\left\|\,\sigma\,\right\|_T\,\left\|\,\rho\,\right\|_T}.
\end{equation}
The cosine of an angle calculated in accordance to
 (\ref{angle}) has the structure of
Pearson's coefficient of linear correlations.
The notion of angle between any two nodes of the
graph arises naturally as soon as we
become interested in
the strength and direction of
a linear relationship between
the flows of random walks moving through them.
If the cosine of an angle (\ref{angle}) is 1
(zero angles),
there is an increasing linear relationship
between the flows of random walks through both nodes.
Otherwise, if it is close to -1 ($\pi$ angle),
  there is
a decreasing linear relationship.
The  correlation is 0 ($\pi/2$ angle)
if the variables are linearly independent.
It is important to mention that
 as usual the correlation between nodes
does not necessary imply a direct causal
relationship (an immediate connection)
between them.

In order to illustrate the approach, we have studied 
five different patterns of compact urban transport networks.
Two of them are
situated on islands: the street network in Manhattan 
and the network of Venetian canals. 
We have also considered two 
medieval German cities developed within the 
fortresses: Rothenburg ob der Tauber in Bavaria 
and the downtown of Bielefeld in Eastern Westphalia.
To supplement the study of urban canal networks, we have 
also examined it 
in the city of Amsterdam. 
In all analyzed transport networks, we observe that the segregation of
a node measured by (\ref{sqaured_norm})
scales negatively with its connectivity;
the slopes of the regression lines slightly exceed 2.

In the present Letter, we describe 
the
Euclidean space structure of space syntax for the spatial network
of 96 Venetian canals which serve the function of roads in the
ancient city that stretches across 122 small islands. While
identifying a canal over the plurality of water routes on the city
map of Venice, the canal-named approach has been used, in which
two different arcs of the city canal network were assigned to the
same identification number provided they have the same name.

In accordance to (\ref{sqaured_norm}), the density ${\bf 1}_i$, which
equals 1 at $i\in V$
and zero otherwise,
acquires the norm $\left\|\,{\bf 1}_i\,\right\|_T$
associated to random walks.
 Its square,
\begin{equation}
\label{norm_node}
\left\|\,{\bf 1}_i\,\right\|_T^2\, =\,\frac 1{\pi_i}\,\sum_{s=2}^N\,
\frac{\,\psi^2_{s,i}\,}{\,1-\mu_s\,},
\end{equation}
expresses
the {\it access time} to a target node in
random walk theory \cite{Lovasz}
quantifying the expected number
of steps
required for a random walker
to reach the node
$i\in V$ starting from an
arbitrary
node  chosen randomly
among all other
nodes  with respect to
the stationary distribution $\pi$.

\begin{figure}[ht]
 \noindent
\begin{center}
\epsfig{file=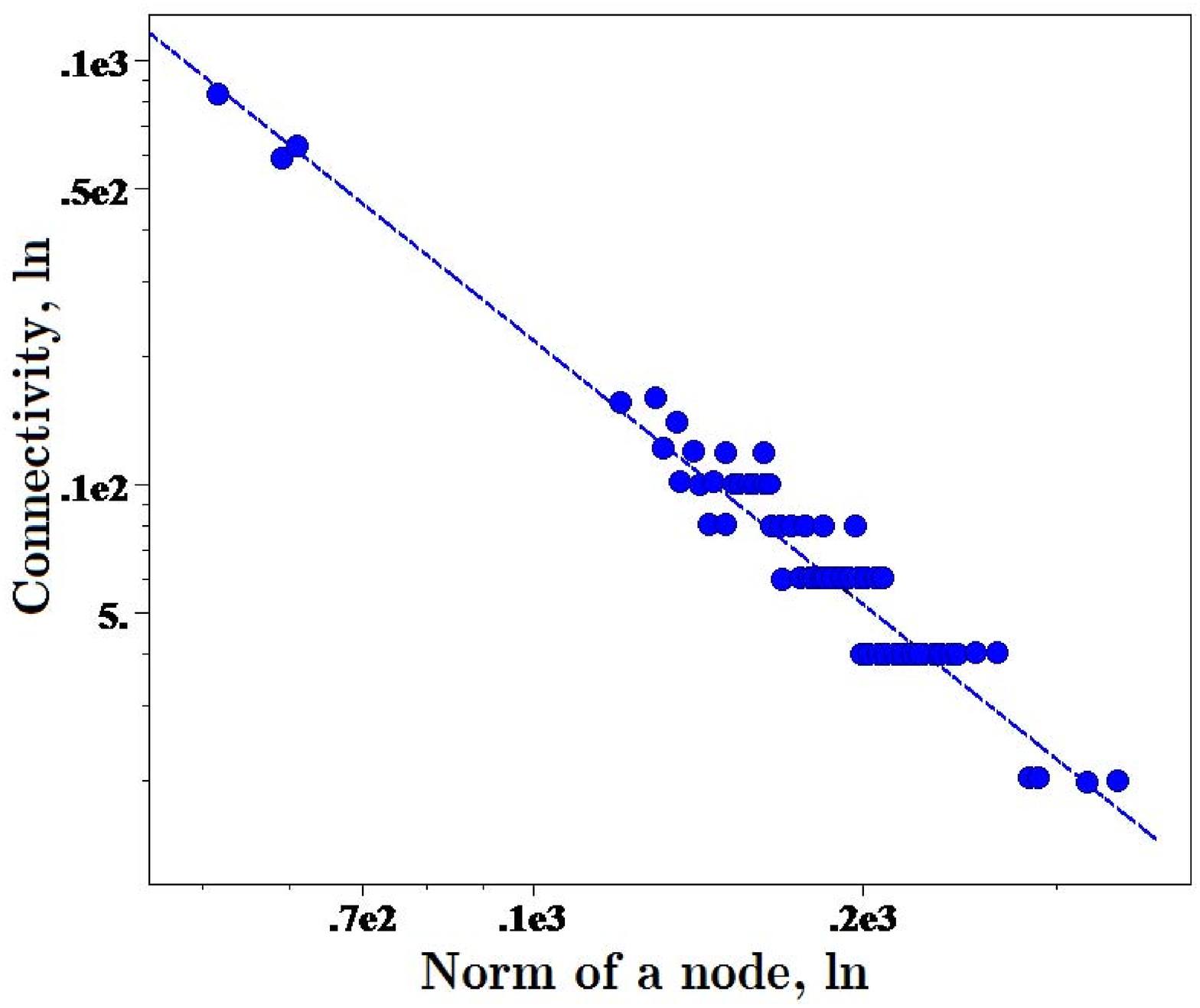,  angle= 0,width =7cm, height =6.5cm}
  \end{center}
\caption{\small The scatter plot of the connectivity vs. the norm a node in
the dual graph representation of 96 Venetian canals acquires with respect
to random walks.
Three data points characterized by the shortest access times represent the
main water routes of Venice: the Lagoon of Venice, the Giudecca canal, and
the Grand canal. Four data points of the worst accessibility are for the
canal subnetwork of Venetian Ghetto.
The slope of the regression
 line equals 2.07.}
\label{Fig2_004a}
\end{figure}

The notion of spatial
segregation acquires a statistical interpretation
with respect to random walks by means of (\ref{norm_node}).
In urban
spatial networks encoded by their dual graphs, the access times
(\ref{norm_node}) strongly
vary  from one open space
to another and could be very large for
statistically segregated spaces.
It is remarkable that the norm a canal of Venice
acquires with respect to random walks
scales with its connectivity (see Fig.~\ref{Fig2_004a}).

The Euclidean distance between ${\bf 1}_i$ and ${\bf 1}_j$
induced by random walks,
\begin{equation}
\label{commute}
\left\|\, {\bf 1}_i-{\bf 1}_j\,\right\|^2_T
\,=\, \sum_{s=2}^N\,\frac 1{1-\mu_s}\left(\frac{\psi_{s,i}}{\sqrt{\pi_i}}-\frac{\psi_{s,j}}{\sqrt{\pi_j}}\right)^2,
\end{equation}
is the {\it commute time} in theory of random walks being
equal to the expected number of steps required for a random
walker starting at $i\,\in\, V$ to visit $j\,\in\, V$ and then to
return back to $i$,  \cite{Lovasz}.

\begin{figure}[ht]
 \noindent
\begin{center}
\epsfig{file=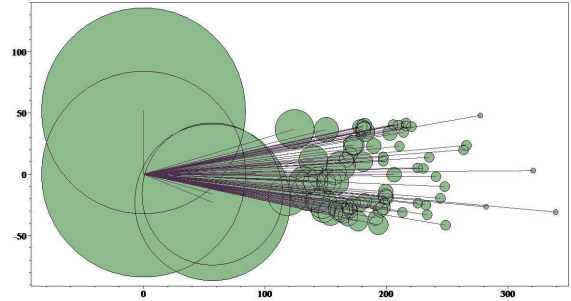,  angle= 0,width =7.5cm, height =4cm}
  \end{center}
\caption{\small The 2-dimensional projection of space syntax of
Venetian canals built from the perspective of the Grand canal of Venice
chosen as the origin. The labels of the
horizontal axes display the expected number of random walk steps.
The labels of the
vertical axes show the degree of nodes (radiuses of the disks).}
\label{Fig2_03bb}
\end{figure}

Indeed, the structure of vector space  $\mathbb{R}^{N-1}$ induced
by random walks cannot be represented visually, however if we
choose a node of the graph as a point of reference, we can draw
the 2-dimensional projection of  Euclidean space by arranging
other nodes at the distances and under the angles they are with
respect to the chosen reference node. The 2-dimensional projection
of the  Euclidean space of Venetian canals  set up by random walks
drawn for the
 the Grand Canal of Venice (the point $(0,0)$) is shown in Fig.~\ref{Fig2_03bb}.
Nodes of the dual graph representation of
 the canal network in Venice
are shown by disks with radiuses taken equal to the degrees of the
nodes. All distances between the chosen origin and other nodes of the
graph have been calculated in accordance to
(\ref{commute}) and (\ref{angle}) has been used in
order to compute angles between nodes.
Canals positively correlated with the Grand Canal of Venice are
set under negative angles (below the horizontal), and under
positive angles (above the horizontal) if otherwise.

The radiuses of disks display the equilibrium configuration of
flows along the Venetian canals  when the traffic volume is conserved.
It is evident from Fig.~\ref{Fig2_03bb}
that disks of smaller radiuses
demonstrate a clear
tendency to be located far
away from the origin
being characterized by
the excessively long commute times with the reference point (the Grand canal of Venice),
while the large disks which stand in Fig.~\ref{Fig2_03bb} for the main water routes
are settled in the closest proximity to the
origin that intends an immediate access to them.

Probably, the most important conclusion of space syntax theory is that
the adequate
 level of the positive relationship between
the connectivity
of city spaces and their integration property (vs. segregation)
called intelligibility
encourages
  peoples way-finding abilities \cite{Hillier:1999}.
Intelligibility of Venetian
  canal network reveals itself quantitatively in the scaling of the
   norms of nodes with connectivity shown in Fig.~\ref{Fig2_004a} and
   qualitatively in the tendency of smaller disks to be located on the
    outskirts of the Venetian space syntax displayed in Fig~\ref{Fig2_03bb}.

The support from the Volkswagen Foundation (Germany) in the
framework of the project "{\it Network formation rules, random set
graphs and generalized epidemic processes}" is gratefully
acknowledged.

\end{document}